# Automated segmentation of retinal fluid volumes from structural and angiographic optical coherence tomography using deep learning

Yukun Guo,[1] Tristan T. Hormel,[1] Honglian Xiong,[1,2] Jie Wang,[1,3] Thomas S. Hwang,[1] and Yali Jia[1,3]

1 Casey Eye Institute, Oregon Health & Science University, Portland, OR 97239, USA
2 School of Physics and Optoelectronic Engineering, Foshan University, Foshan, Guangdong 528000, China
3 Department of Biomedical Engineering, Oregon Health & Science University, Portland, OR 97239, USA

**Address for correspondence**:

Yali Jia, Ph.D.

Casey Eye Institute

Oregon Health & Science University

Portland, OR 97239, USA

Email: jiaya@ohsu.edu

**Word count**: 5133

**Funding Information**:

National Institutes of Health (R01 EY027833, R01 EY024544, P30 EY010572);

Unrestricted Departmental Funding Grant and William & Mary Greve Special Scholar Award from Research to Prevent Blindness (New York, NY).

**Commercial Relationships Disclosure**:

**Yukun Guo,** None; **Tristan T. Hormel,** None; **Honglian Xiong,** None; **Jie Wang,** None; **Thomas S. Hwang,** None; **Yali Jia**, Optovue (F, P)

**ABSTRACT**

**Purpose**: We proposed a deep convolutional neural network (CNN), named Retinal Fluid Segmentation Network (ReF-Net) to segment volumetric retinal fluid on optical coherence tomography (OCT) volume.

**Methods**: 3 × 3-mm OCT scans were acquired on one eye by a 70-kHz OCT commercial AngioVue system (RTVue-XR; Optovue, Inc.) from 51 participants in a clinical diabetic retinopathy (DR) study (45 with retinal edema and 6 healthy controls). A CNN with U-Net-like architecture was constructed to detect and segment the retinal fluid. Cross-sectional OCT and angiography (OCTA) scans were used for training and testing ReF-Net. The effect of including OCTA data for retinal fluid segmentation was investigated in this study. Volumetric retinal fluid can be constructed using the output of ReF-Net. Area-under-Receiver-Operating-Characteristic-curve (AROC), intersection-over-union (IoU), and F1-score were calculated to evaluate the performance of ReF-Net.

**Results**: ReF-Net shows high accuracy (F1 = 0.864±0.084) in retinal fluid segmentation. The performance can be further improved (F1 = 0.892±0.038) by including information from both OCTA and structural OCT. ReF-Net also shows strong robustness to shadow artifacts. Volumetric retinal fluid can provide more comprehensive information than the 2D area, whether cross-sectional or *en face* projections.

**Conclusions**: A deep-learning-based method can accurately segment retinal fluid volumetrically on OCT/OCTA scans with strong robustness to shadow artifacts. OCTA data can improve retinal fluid segmentation. Volumetric representations of retinal fluid are superior to 2D projections.

**Translational Relevance**: Using a deep learning method to segment retinal fluid volumetrically has the potential to improve the diagnostic accuracy of diabetic macular edema by OCT systems.

# 1. INTRODUCTION

Diabetic macular edema (DME) is the most common cause of vision loss in diabetic retinopathy (DR) [1]. Accurate detection of DME for screening and treatment response is critical in preventing vision loss [2]. Currently, clinicians use structural optical coherence tomography (OCT) to diagnose DME from retinal thickness maps, central macular thickness (CMT), and qualitative inspection of the raster scans [3]. CMT is an imperfect biomarker for DME, as atrophy due to cell loss can reduce the thickness in the presence of edema, and the presence of other pathology such as epiretinal membranes can increase the thickness without edema [4,5]. Segmentation and quantification of retinal fluid cysts provide a more specific biomarker for DME that is useful even when other confounding abnormalities are present [6].

Several reading centers have explored the measurement of retinal fluid area on cross-sectional commercially available OCT. However, quantification of fluid based on the quantity of fluid present may not be accurately captured by cross-sections at large step intervals. Previously, we presented a fuzzy level-set method [7] to measure retinal fluid volume in OCT/OCT angiography (OCTA) scans of DME eyes. This approach, however, was vulnerable to the shadow artifacts caused by vitreous floaters and large vessels as well as pupil vignetting.

Contemporary deep-learning-based methods have shown a great advantage for image segmentation tasks [8–15]. In ophthalmology, researchers have proposed a number of deep neural networks to solve specific problems, such as retinal layer segmentation in OCT [16–18], retinal vessel segmentation in fundus photography [19,20], and retinal nonperfusion area segmentation in OCTA [21–23]. Deep-learning-based retinal fluid segmentation on cross-sectional OCT has also been reported by many scholars. Bai et.al. used a fully convolutional neural network (CNN) and a fully connected conditional random field method to segment cystoid macular edema [24]. This method can get good segmentation results when the fluid deposits are extensive, but is not sensitive to small target regions. Schlegl et. al. proposed an encoder-decoder-based deep learning method to detect and quantify macular fluid in OCT images that achieved high accuracy [25]. In order to solve the challenges due to speckle noise and imaging artifacts, Girish et al. added denoising and sub-retinal layer segmentation during pre-processing before feeding the data to a CNN, which improved performance [26]. Denoising also helped their algorithm perform well on data from different instruments. Li et al. [27] applied a 3D CNN on Spectralis OCT (Heidelberg Engineering Inc.) scans and achieved high performance, but the sparse sampling density hindered the accurate measurement of fluid volume. However, all of these methods segment retinal fluid from OCT data alone. We hypothesize that OCTA signal could improve segmentation accuracy, as retinal fluid and blood flow are never collocal.

In this study, we propose a new deep CNN, named Retinal Fluid Segmentation Network (ReF-Net), to segment intraretinal and subretinal fluid from simultaneously generated volumetric OCT/OCTA scans. Our network provides three key innovations: first, we include OCTA data in the network input. As part of this work, we characterized the effect of this inclusion on network performance. Second, we provide 3D segmentation results and data representations. Last, our network shows strong, robust performance with different types of shadow artifacts.

## 2. METHODS

### 2.1 Data acquisition

Volumetric OCT data were acquired over the central 3×3-mm region using a 70-kHz OCT commercial AngioVue system (RTVue-XR; Optovue, Inc.) centered at 840nm with a full-width half-maximum bandwidth of 45nm. Two repeated B-scans were taken at each 304 raster positions, and each B-scan consists of 304 A-lines. The structural OCT was generated by averaging the two repeated B-scans, and the OCTA was generated by using the split-spectrum amplitude-decorrelation angiography (SSADA) algorithm [28] to compute the decorrelation between the two repeated B-scans, simultaneously. Projection-resolved OCTA (PR-OCTA) removed shadow graphic artifacts from the superficial vasculatures while preserving true flow signal in deeper layers [29].

### 2.2 Convolutional neural network architecture

Deep convolutional neural networks (CNNs) are superior to traditional methods for semantic segmentation tasks. U-Net-like CNNs have high adaptability to medical image segmentation due to skip connections that enable feature extraction with minimal loss of resolution [30]. In this study, we adopted U-Net-like architecture and designed ReF-Net to segment retinal fluid. In order to increase its capability for feature extraction, we applied some modifications to the original U-Net [Fig.1 (A)]. A Multi-scale feature extraction block [Fig.1(B)], inspired by Inception [31], is placed after the input layer. This block can extract multi-scale features that enhance the ability of the neural network to detect targets with different sizes[21,22]. We also replaced the regular forward convolutional layers with residual units [Fig. 1 (C-D)], borrowed from ResNet [32], to increase the feature extraction ability of ReF-Net.

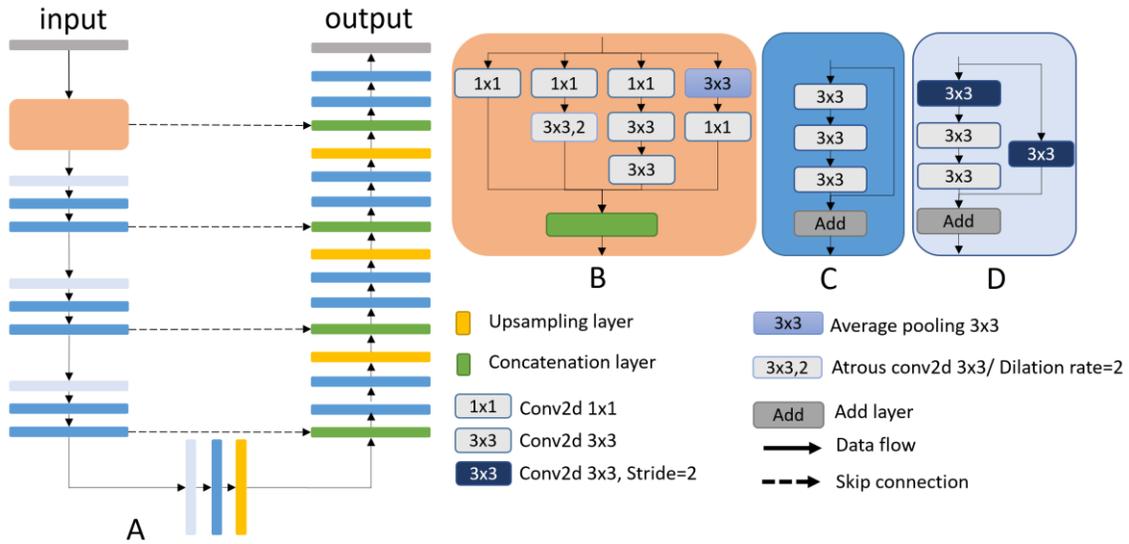

Fig. 1. The architecture of deep convolutional neural network constructed in this study. (A) ReF-Net architecture. (B) Multi-scale block. (C-D) Residual convolutional blocks.

## 2.3 Dataset preprocessing

The data we used in our experiment contains three types of image: structural OCT B-scans [Fig. 2(A)], OCTA B-scans [Fig. 2(B)], and the ground truth map [Fig. 2(C)]. We collected data from a total of 51 eyes (45 with DME and 6 healthy controls) from a clinical DR study, and each eye has two repeated volumetric scans. To remove excessive speckle noise, each B-scan was enhanced by performing a moving average with the two adjacent B-scans. In this study, each volumetric scan contained a total of 304 B-scans.

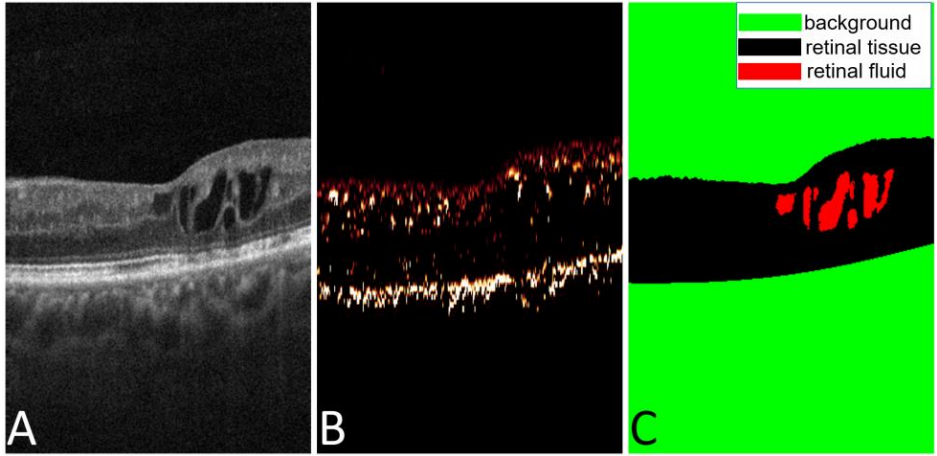

Fig. 2. Representative OCT/OCTA B-scan showing retinal fluid. (A) OCT B-scan. (B) OCTA B-scan. (C) Ground truth map with three categories, background (green), retinal tissue (black), and retinal fluid area (red).

Previously, researchers only used OCT data to segment retinal fluid. As the fluid region does not contain any vasculature, the simultaneously computed OCTA data may contribute to segmentation accuracy. In order to verify that OCTA data can indeed improve the segmentation performance, we designed two versions of the CNN in this study, each of which works with different inputs. The input of the first network (ReF-Net-OCT) only contains OCT data, and the other one (ReF-Net-OCTA) contains both OCT and OCTA data. Before feeding OCT and OCTA data to ReF-Net-OCTA, an image fusion operation [Eq. 1] was applied to merge these two types of data together:

$$I_{\text{fusion}} = (1-\beta) \times I_{\text{OCT}} + \beta \times I_{\text{OCTA}} \qquad (1)$$

Here, $I_{\text{fusion}}$ is the fused data, $\beta \in [0,1]$ is a fusion factor, and $I_{\text{OCT}}$ and $I_{\text{OCTA}}$ represent the OCT and OCTA data, respectively. In order to get the optimal parameter value of $\beta$, we tested 12 values from 0.05 to 0.60 at intervals of 0.05.

The ground truth map that was used to train ReF-Net contains three categories: background, retinal tissue, and retinal fluid area. In order to obtain the ground truth, three graders manually delineated retinal fluid area using a customized graphical user interface (Fig. 3A). A guided bidirectional graph search (GB-GS) method was employed to segment retinal tissue boundaries [33]. We merged three manual grading outputs together by using a pixel-wise voting method to obtain the final ground truth map. [Fig. 3(B)]. In rare instances when three graders assigned a pixel to three disparate categories (1 vote for background, 1 vote for tissue, 1 vote for fluid), the graders would reach a consensus through discussion.

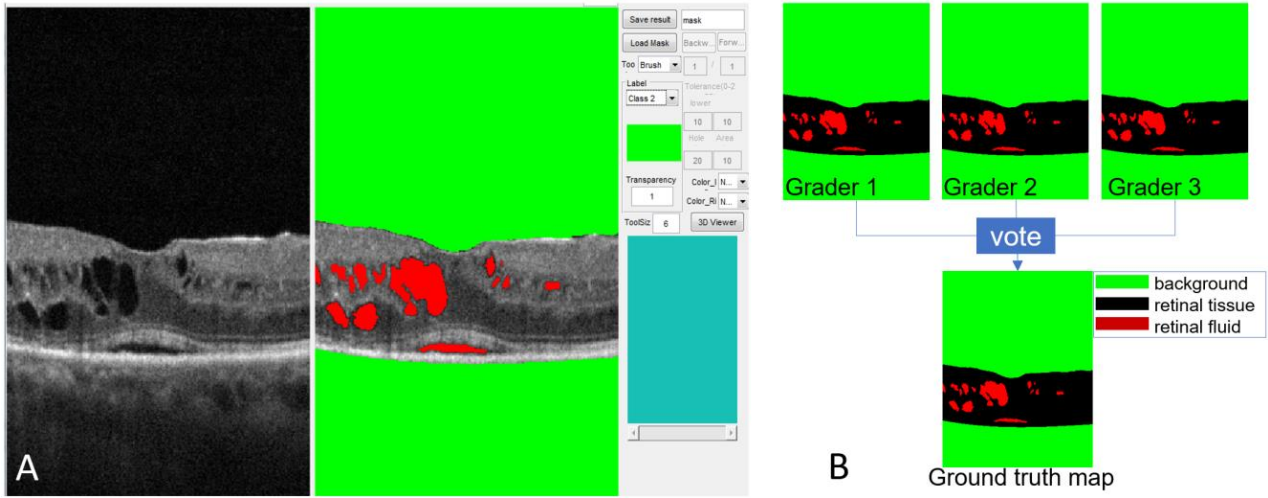

Fig. 3. Manual delineation of the ground truth for training. (A) The in-house graphical user interface software. (B) Three graders manually delineated the background (green), retinal tissue (black), and retinal fluid area (red). Pixel-wise voting method to generate the final ground truth map.

**2.4 ReF-Net hyperparameter settings**

**Loss function**

As a multi-class segmentation task, retinal fluid segmentation encounters a serious category imbalance problem due to the huge difference in the area between these three categories. To suppress the effect of the category imbalance, we used a categorical cross-entropy loss combined with weighted Jaccard coefficient loss [22]:

$$L = \sum_{i=1}^{N} J_i \times w_i, \quad \sum_{i=1}^{N} w_i = 1$$
$$J = \left(1 - \frac{\sum_x y(x) \times \hat{y}(x) + \alpha}{\sum_x (y(x) + \hat{y}(x)) - \sum_x y(x) \times \hat{y}(x) + \alpha}\right) \times \alpha \quad (2)$$

where $N$ is the number of categories, $J_i$ is the Jaccard loss of $i$-th category, and $w_i$ is the weight of $i$-th category associated with Jaccard coefficient $J_i$. In our experiment, we associated the three categories (retinal fluid area, retinal tissue, and background) with weights $w$ = (0.5, 0.25, 0.25). We set a higher value to the retinal fluid region to make the ReF-Net pay more attention to this category. $y$ and $\hat{y}$ denote the ground truth and output of the ReF-Net, respectively, $x$ is the position of each pixel in the sample, and $\alpha$ is a smoothing factor usually set to 100 [22].

**Optimizer and training**

A modified Adam algorithm, AdamW [34], was used to train ReF-Net by minimizing the weighted Jaccard coefficient loss. The initial learning rate was set to 0.001. We used a global learning decay strategy that reduces the learning rate by 90 percent when the loss reaches a plateau. Early stopping was employed to stop the training phase when the loss didn't show a decrease in 15 training epochs. The total dataset (51 eyes) was randomly split into a training set (40 eyes) and test set (11 eyes). The training set was augmented by flipping each sample horizontally.

## 3. RESULTS

ReF-Net was implemented in Python 3.7 with Keras (Tensorflow-backend) on a PC with an Intel i7 CPU, two GTX 1080Ti GPUs, and 64GB RAM.

### 3.1 Segmentation accuracy

The performance of each model (Table. 1) was measured for the Area-under-Receiver-Operating-Characteristic-curve (AROC), intersection-over-union (IoU; also known as Jaccard coefficient), and F1-score, defined as

$$F1 = \frac{2 \times TP}{2 \times TP + FP + FN} \quad (3)$$

where TP is true positive, FP is false positive, FN is false negative. ReF-Net-OCT can achieve good results using only the OCT data as the input, which is expected because the retinal fluid region shows extremely low OCT reflectance intensity compared to the healthy retina. For ReF-Net-OCTA, the AROC, IoU and F1-score depend on the $\beta$ parameter in Eq. 1, with each reaching a peak when $\beta = 0.20$. The value of $\beta$ regulates the proportion of features contributed to segmentation in both OCT and OCTA. As $\beta$ increases, so does the number of OCTA features used for decision making, while the corresponding number of features from OCT decrease. The numbers of effective features achieved a peak when $\beta = 0.20$; thus, we can confirm that the features from the OCT data played the major role in the segmentation, although OCTA still improved performance. Comparing the segmentation results of ReF-Net-OCTA [Fig. 4B] to the ground truth maps [Fig. 4 C], the large overlapping areas [Fig. 4D] indicates the high accuracy of ReF-Net.

Table 1. Agreement (in pixels) between automated detection and manual delineation of retinal fluid region (mean ± standard deviation). Bold indicates best performance.

| CNNs | AROC | IoU | F1-score |
| --- | --- | --- | --- |
| **ReF-Net-OCT** | 0.992±0.011 | 0.770±0.124 | 0.864±0.084 |
| **ReF-Net-OCTA ($\beta$ =0.05)** | 0.995±0.004 | 0.775±0.107 | 0.870±0.073 |
| **ReF-Net-OCTA ($\beta$ =0.10)** | 0.996±0.005 | 0.787±0.124 | 0.875±0.086 |
| **ReF-Net-OCTA ($\beta$ =0.15)** | 0.996±0.007 | 0.800±0.096 | 0.885±0.064 |
| **ReF-Net-OCTA ($\beta$ =0.20)** | **0.996±0.003** | **0.807±0.061** | **0.892±0.038** |
| **ReF-Net-OCTA ($\beta$ =0.25)** | 0.992±0.008 | 0.771±0.089 | 0.868±0.060 |
| **ReF-Net-OCTA ($\beta$ =0.30)** | 0.992±0.007 | 0.760±0.083 | 0.860±0.057 |
| **ReF-Net-OCTA ($\beta$ =0.35)** | 0.991±0.009 | 0.728±0.097 | 0.840±0.065 |
| **ReF-Net-OCTA ($\beta$ =0.40)** | 0.993±0.006 | 0.716±0.100 | 0.830±0.069 |
| **ReF-Net-OCTA ($\beta$ =0.45)** | 0.992±0.013 | 0.672±0.128 | 0.797±0.091 |
| **ReF-Net-OCTA ($\beta$ =0.50)** | 0.989±0.015 | 0.664±0.169 | 0.784±0.137 |
| **ReF-Net-OCTA ($\beta$ =0.55)** | 0.988±0.013 | 0.636±0.195 | 0.757±0.172 |
| **ReF-Net-OCTA ($\beta$ =0.60)** | 0.988±0.017 | 0.638±0.190 | 0.760±0.162 |

### 3.2 Resistance to shadow artifacts

Shadow artifacts caused by large vessels, vitreous floaters, and pupil vignetting can reduce the signal reflectance strength in retinal tissue, which reduces contrast in shadow area. To verify the robustness of ReF-Net on shadow artifacts, we applied ReF-Net (ReF-Net-OCTA, $\beta$ =0.20) on cases with three types of typical shadow artifacts [Fig. 5]; ReF-Net could handle all three types. This is likely because ReF-Net doesn't just rely on contrast information between the fluid and tissue, but also on other geometrical information extracted by convolutional kernels.

### 3.3 Volumetric versus 2D projected retinal fluid and cross-sectional quantification

By applying the best model (ReF-Net-OCTA, $\beta$ =0.20) to each frame of OCT/OCTA data, we can construct a 3D segmentation. Compared to the 2D *en face* projected retinal fluid regions overlaid on projected OCT reflectance [Fig. 6 (A1-D1)], the 3D retinal fluid region [Fig. 6 (A2-D2)] is more intuitive. Furthermore, the 3D volumetric result produces a more meaningful quantification than the 2D projected area, which may not reflect the actual extent of the fluid region. In [Fig. 6 (A1, B1)], for example, although the two DME cases have similar fluid areas in the 2D *en face* image, the actual fluid volumes differ by a factor of 2 [Fig.

6 (A2, B2)]. Similarly, cases with very different 2D projected fluid area on *en face* image [Fig. 6 (C1, D1)], may have similar fluid volume [Fig. 6 (C2, D2)].

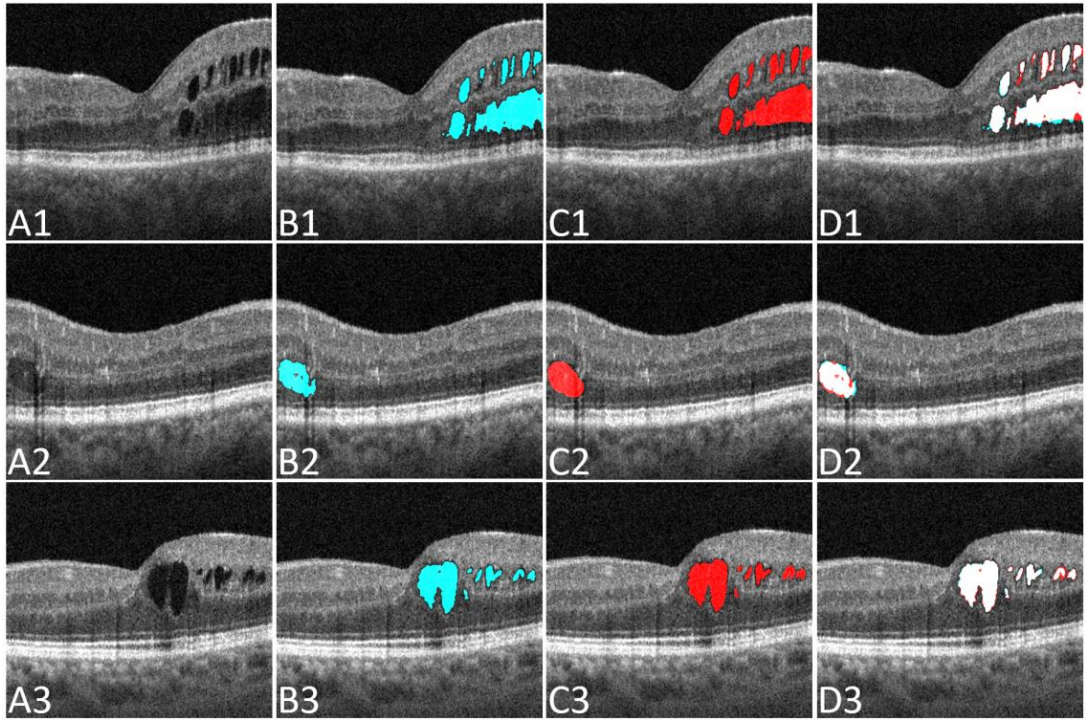

Fig. 4 Comparison between ReF-Net-OCTA (β =0.20) and ground truth on structural OCT B-scans. (A) Structural OCT B-scans. (B) Segmented fluid maps from ReF-Net (blue) and (C) the ground truth maps (red) overlaid on structural cross-sections. (D) Difference map between segmented fluid from ReF-Net and ground truth. White area is the overlap region of two maps. The blue and red in (D) show pixels exclusively in the algorithm output or ground truth, respectively.

Figure 7 further demonstrates how 2D fluid measurement on cross-sectional raster scans are more likely to miss a substantial portion of retinal fluid [Fig. 7] due to undersampling. Using densely sampled OCT and OCTA data, ReF-Net can render retinal fluid cysts in 3 dimension (the volumetric of fluid is shown in Fig. 7B, blue) and reveal anatomic changes related to macular edema more completely.

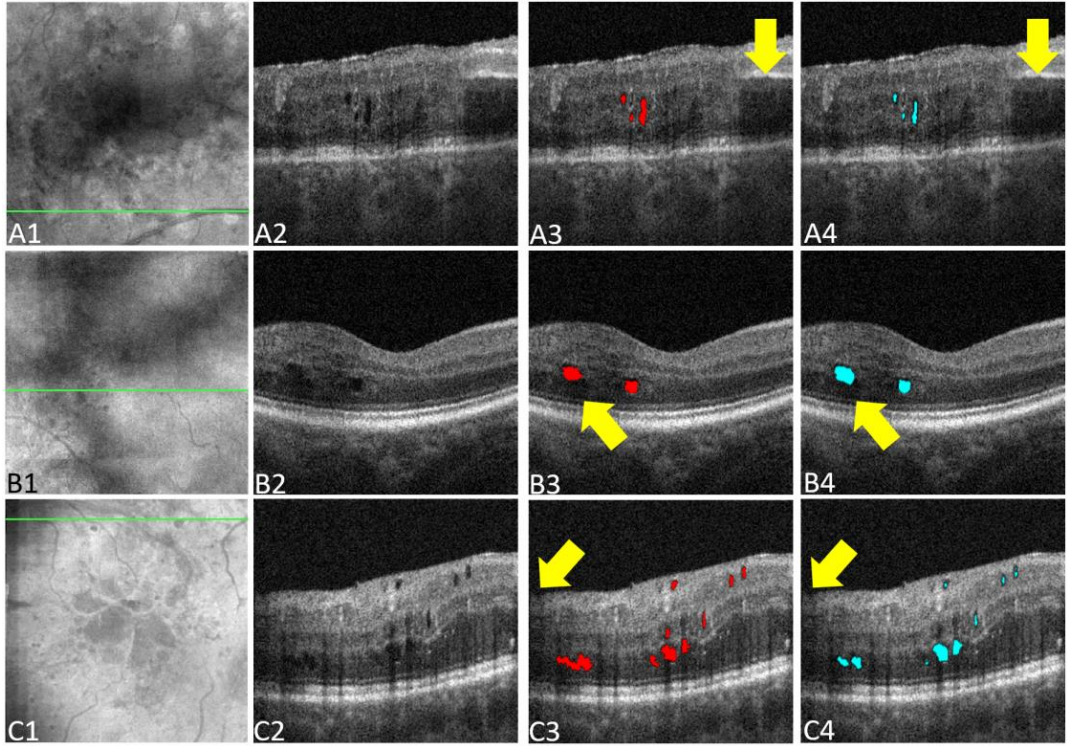

Fig. 5 Automated retinal fluid segmentation results on shadow artifact effected scans. Yellow arrows indicate the shadow artifacts. (Row A) Example case with large vessel shadow artifacts. (Row B) Example case with vitreous floater shadow artifacts. (Row C) Example case with pupil vignetting shadow artifacts. (Column 1) Reflectance *en face* images, with the green line indicating the position of the B-scan shown in the other columns. (Column 2) Raw cross-sectional scans. (Column 3) Ground truth map (red) overlaid on B-scans. (Column 4) ReF-Net outputs (blue) overlaid on the B-scans.

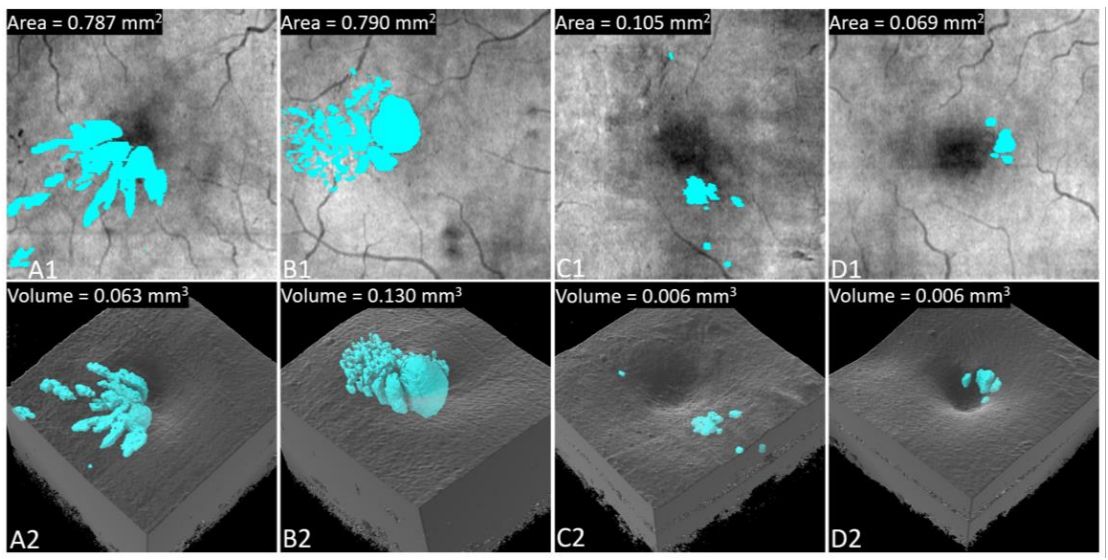

Fig. 6 Comparison between 2D projected fluid areas and 3D fluid volumes in DME cases. (A1-D1) 2D structural OCT and retinal fluid projections. (A2-D2) 3D structural OCT and retinal fluid representations. Apparent fluid areas can be similar while volumes are quite different (A, B), and apparent fluid areas can be quite different while volumes are similar (C, D). In such cases, the 2D projection is misleading.

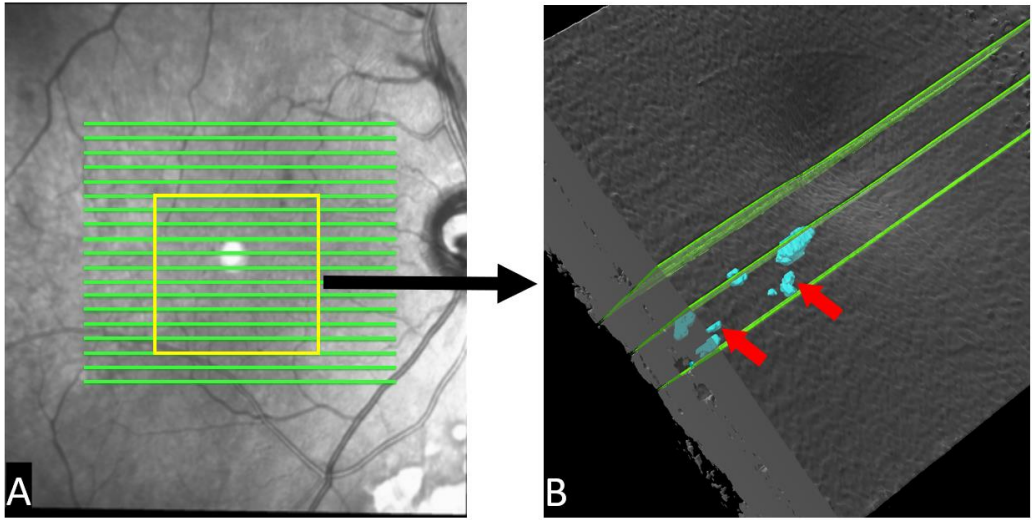

Fig. 7 A DME case with substantial portion of retinal fluid missed by under sampled structural OCT. (A) Infrared photograph with sampling positions (green lines) from a Spectralis OCT (Heidelberg Engineering Inc.) scan. (B) Volumetric OCT with retinal fluid volume (blue). Red arrows indicate the missed retina fluid area; green lines indicate the sampling position.

### 3.4 Recovering DME diagnosis from a false-negative CMT measurement.

Central macular thickness (CMT) is a commonly used biomarker for DME diagnosis. Major clinical trials have used CMT greater than 2 standard deviations from the population mean for inclusion criteria in clinical trials [35]. Because there is a significant population variation in retinal thickness, and specific pathology such as epiretinal membrane or retinal atrophy can cause changes unrelated to macular edema, diagnosis of DME solely based on CMT measurements can be unreliable. Fig. 8 shows an eye with a CMT of 217 μm, which does not meet the CMT definition of center-involved DME, but has a known clinical diagnosis of DME with retinal fluid caused by DME [3]. ReF-Net automatically detected a fluid volume of 0.044 mm$^3$.

### 3.5 Longitudinal study of retinal fluid in OCT/OCTA scans

The change in retinal fluid volume is an important indicator of treatment response in DME. Using our algorithm, it is easy to visualize fluid volume changes longitudinally. To do so, we register the baseline and follow up scans [Fig. 9 (A, B)] using Bruch's membrane and large retinal vessels as a reference for the axial and lateral directions, respectively. After the omnidirectional registration [Fig. 9 C], the changes in the shape and size of the fluid [Fig. 9 D] can be easily visualized. Furthermore, we can identify the vascular changes caused by retinal fluid accumulation by overlaying the retinal fluid volumes on the angiographic volumes [Fig. 9 (E, F)].

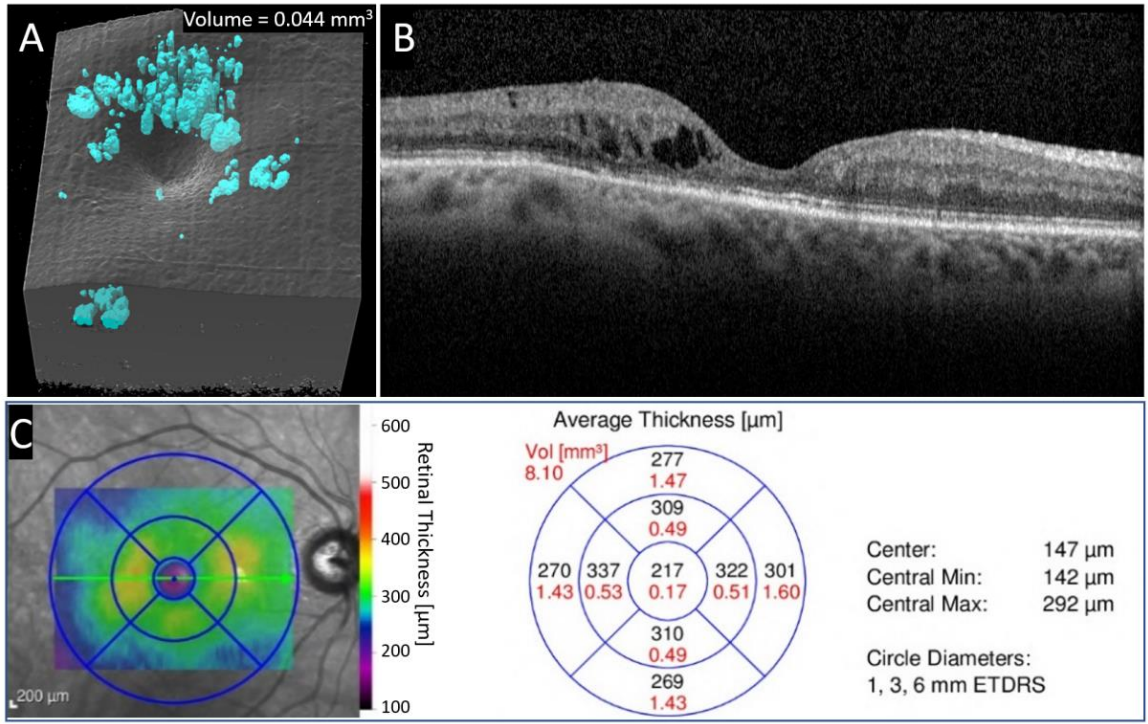

Fig. 8 A diabetic macular edema (DME) case with a false negative result from central macular thickness (CMT) was automatically detected and measured by ReF-Net. (A) Retinal fluid volume segmented by ReF-Net. (B) Cross-sectional structural OCT. (C) Retinal thickness map and average thickness distribution in early treatment diabetic retinopathy study (ETDRS) grid. The CMT value is 217, which does not meet the definition of DME.

## 4. DISCUSSION

We have presented a deep-learning-based method (ReF-Net) for segmenting and quantifying retinal fluid in 3 dimensions using OCT and OCTA volumes. We demonstrated that OCTA data enhances the segmentation task, and the 3D approach provides a more intuitive and complete representation of the anatomic changes in DME than 2D cross-sections or projections.

ReF-Net is a U-Net-like convolutional neural network. We added several useful modifications to enhance its feature extraction capability, such as a multi-scale feature extraction block and residual blocks. Comparing to previous deep-learning-based methods, ReF-Net achieved high accuracy [Table. 2]. The methods that proposed by Bai et al. [24] and Schlegl et al. [25] (both of which were based on fully convolutional neural networks) shown lower performance than U-Net-like CNNs (the other four methods in Table 2). This may because the skip connections could transfer feature information from the initial layers to the deeper layers directly, thereby improving the CNN's ability to identify minute details in the target. Comparing to Girish's [26] and Lu's [36] methods, ReF-Net shows higher accuracy, which is likely because the multi-scale feature extraction block and residual blocks made ReF-Net more adaptable to different sizes of edema. Although Li's method [27] achieved high accuracy on sparsely sampled OCT volumes, the method may not reliably quantify volumes due to the low sampling rate that can easily miss small fluid deposits [Fig. 8]. Even if all the volumes are detected, a low sampling density will cause the precision of the volume measurement to suffer.

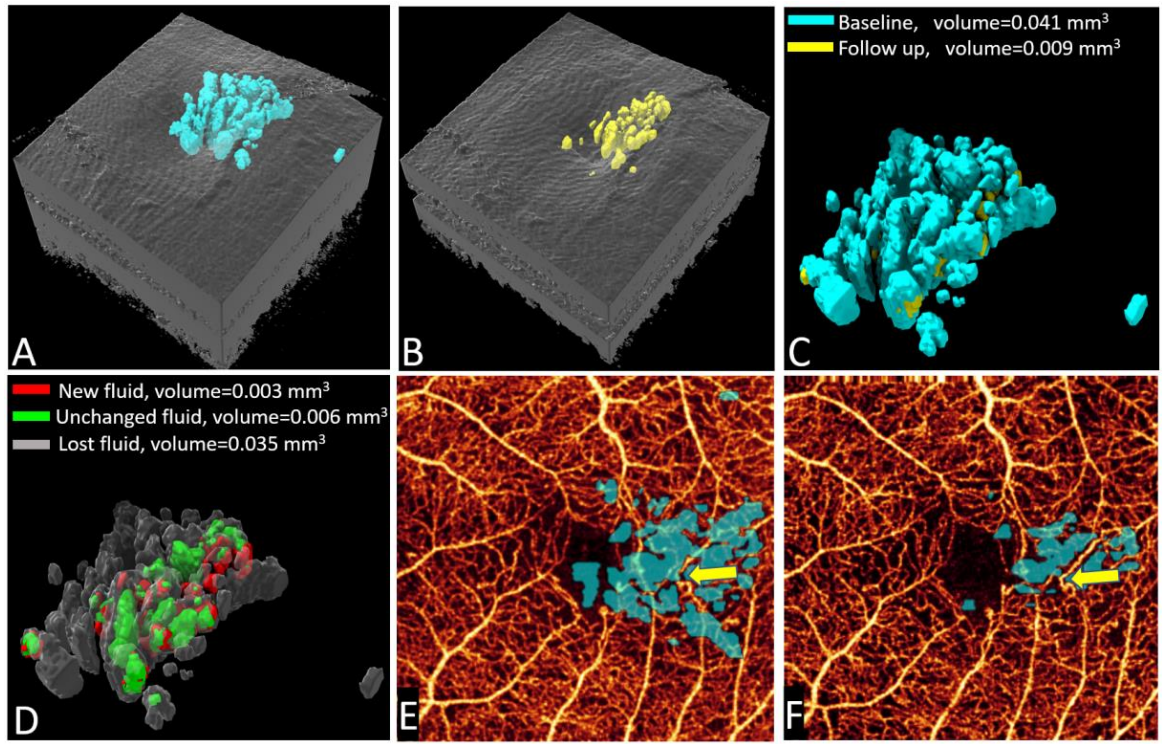

Fig. 9 Local dynamics of retinal fluid in longitudinal monitoring of a DME eye. (A) Baseline. (B) One year follow up after the treatment. (C) Registered baseline and follow up scans. (D) Changes in the retinal fluid region. (E) Baseline retinal fluid area overlaid on an inner retinal OCT angiogram. (F) Follow up retinal fluid area overlaid on an inner retinal OCT angiogram. The yellow arrow indicated the change of vasculature caused by retinal fluid.

Table 2. Performance comparison between deep-learning-based methods (mean ± standard deviation). Some uncertainties and metrics were not reported in the original studies.

| Methods | F1-Score (Dice coefficient) | AROC |
|---|---|---|
| Bai et al.[24] | 0.610±0.210 | - |
| Schlegl et al.[25] | - | 0.940 |
| Girish et al.[26] | 0.720±0.190 | - |
| Li et al.[27] | 0.955 | - |
| Lu et al.[36] | 0.767 | 1.000 |
| ReF-Net (proposed) | 0.892±0.038 | 0.996 |

The main feature useful for retinal fluid segmentation is the low reflectance of fluid regions in structural OCT scans. Additionally, since the fluid region does not contain any flow signal, OCTA data have the potential to improve segmentation performance. In this study, we analyzed the contribution of OCTA data to ReF-Net-OCTA's performance. By merging OCT and OCTA data with different fusion factors, we trained 12 different models based on the ReF-Net-OCTA architecture, each with different segmentation accuracy. Experimental results show that ReF-Net-OCTA achieves the best performance (F1-score>0.89) when the fusion factor equals 0.2 (Table. 1). The accuracy of ReF-Net-OCTA is superior to ReF-Net-OCT for this value of $\beta$, indicating that some OCTA data is helpful to segment fluid correctly. Additional benefits of using OCT/OCTA data are: (1) high-sampling-density structural OCT and OCTA can be simultaneously processed from the same scan; (2) they are inherently co-registered, facilitating the study on anatomic and angiographic pathologies; (3) OCTA vasculatures can be used for registering multiple scans during longitudinal studies.

Since the input to ReF-Net is a single B-scan, ReF-Net is compatible with conventional cross-sectional data. Additionally, ReF-Net also achieves high performance on scan volumes with structural OCT only, which allows us to apply ReF-Net to the scans that only have OCT data at the expense of some accuracy loss. Furthermore, ReF-Net-OCT/OCTA does not require segmented layers for its input. This is a critical advantage in this context, since retinal slab segmentation is especially error prone in the presence of pathological disruptions to normal slab anatomy, of which the presence of retinal fluid is an important example. ReF-Net also shows strong robustness on different types of topical shadow artifacts, including large vessel shadows, vitreous floater shadows, and pupil vignetting shadows. ReF-Net was trained on 3×3-

mm central macular scans, but it could also easily be migrated to larger scan patterns using transfer learning [37].

Although our method can segment retinal fluid with high accuracy, there are some drawbacks that may hinder its application. Retinal fluid can be classified into intraretinal fluid and subretinal fluid according to location, with each category having different diagnostic and prognostic value [38]. Because a CNN is insensitive to the location of the target and the different types of retinal fluid have similar features, we labeled all the fluid regions in one category to help ReF-Net learn consistent features in order to improve its performance. Thus, ReF-Net cannot distinguish these two types of retinal fluid. However, differentiating type could be easily accomplished by an additional retinal layer segmentation step. Alternatively, other unidentified fluid accumulation features that could be discovered with the precise 3D segmentation provided by ReF-Net could possibly serve as superior biomarkers than the intra/subretinal fluid classification. Another limitation concerns the features used for decision making. The most obvious feature indicating retinal fluid is the high contrast between the fluid region and retinal tissue. In future work, we can try to improve the performance of ReF-Net by resolving this drawback.

## 5. Conclusions

In summary, we designed a deep-learning-based method named ReF-Net to segment volumetric retinal fluid on OCT/OCTA scans. By combining the OCT and OCTA data as the input to the network, ReF-Net demonstrated that OCTA data can improve retinal fluid segmentation. Our results indicate volumetric representations of retinal fluid can provide more comprehensive information than 2D either cross-sections or projections.

**ACKNOWLEDGMENT**

This work was supported by grant National Institutes of Health (R01 EY027833, R01 EY024544, P30 EY010572); Unrestricted Departmental Funding Grant and William & Mary Greve Special Scholar Award from Research to Prevent Blindness (New York, NY).